%% file: main.tex
\documentclass[conference]{IEEEtran}

\usepackage{tikz}
\usepackage{amsmath}
\usepackage{url}
\usepackage{balance}
\usepackage{listings}
\usepackage{calc}
\usepackage{bbding}
\usepackage{pifont}
\usepackage{wasysym}
\usepackage{amssymb}
\usepackage{xcolor}
\usepackage{graphicx}
\usepackage{caption,subcaption}
\usepackage{enumitem}
\usepackage{supertabular}
\usepackage{microtype}
\usepackage{fancyvrb}
\usepackage{multicol}
\usepackage{rotating}
\usepackage{enumitem}
\usepackage{diagbox}
\usepackage{xspace}
\usepackage{hyperref}
\usepackage[ruled,vlined,linesnumbered]{algorithm2e}
\usepackage{tikz}
\usepackage{mdframed}
\usepackage{multirow}
\usepackage{booktabs}
\usepackage{mathtools}
\usepackage{backnaur}
\usepackage{comment}
\usepackage{tcolorbox}
\usepackage{lmodern}
\usepackage{siunitx}
\usepackage{etoolbox}

\renewrobustcmd{\bfseries}{\fontseries{b}\selectfont}
\renewrobustcmd{\boldmath}{}
\newrobustcmd{\B}{\bfseries}
\definecolor{lightgray}{rgb}{.9,.9,.9}
\definecolor{darkgray}{rgb}{.4,.4,.4}
\definecolor{purple}{rgb}{0.65, 0.12, 0.82}
\definecolor{pinegreen}{rgb}{0.0, 0.47, 0.44}

\definecolor{verylightgray}{rgb}{.97,.97,.97}

\lstdefinelanguage{Solidity}{
	keywords=[1]{anonymous, assembly, assert, balance, break, call, callcode, case, catch, class, constant, continue, constructor, contract, debugger, default, delegatecall, delete, do, else, emit, event, experimental, export, external, false, finally, for, function, gas, if, implements, import, in, indexed, instanceof, interface, internal, is, length, library, log0, log1, log2, log3, log4, memory, modifier, new, payable, pragma, private, protected, public, pure, push, require, return, returns, revert, selfdestruct, send, solidity, storage, struct, suicide, super, switch, then, this, throw, transfer, true, try, typeof, using, view, while, with, addmod, ecrecover, keccak256, mulmod, ripemd160, sha256, sha3}, 
	keywordstyle=[1]\color{blue}\bfseries,
	keywords=[2]{address, bool, byte, bytes, bytes1, bytes2, bytes3, bytes4, bytes5, bytes6, bytes7, bytes8, bytes9, bytes10, bytes11, bytes12, bytes13, bytes14, bytes15, bytes16, bytes17, bytes18, bytes19, bytes20, bytes21, bytes22, bytes23, bytes24, bytes25, bytes26, bytes27, bytes28, bytes29, bytes30, bytes31, bytes32, enum, int, int8, int16, int24, int32, int40, int48, int56, int64, int72, int80, int88, int96, int104, int112, int120, int128, int136, int144, int152, int160, int168, int176, int184, int192, int200, int208, int216, int224, int232, int240, int248, int256, mapping, string, uint, uint8, uint16, uint24, uint32, uint40, uint48, uint56, uint64, uint72, uint80, uint88, uint96, uint104, uint112, uint120, uint128, uint136, uint144, uint152, uint160, uint168, uint176, uint184, uint192, uint200, uint208, uint216, uint224, uint232, uint240, uint248, uint256, var, void, ether, finney, szabo, wei, days, hours, minutes, seconds, weeks, years},	
	keywordstyle=[2]\color{teal}\bfseries,
	keywords=[3]{block, blockhash, coinbase, difficulty, gaslimit, number, timestamp, msg, data, gas, sender, sig, value, now, tx, gasprice, origin},	
	keywordstyle=[3]\color{violet}\bfseries,
	identifierstyle=\color{black},
	sensitive=false,
	comment=[l]{//},
	morecomment=[s]{/*}{*/},
	commentstyle=\color{gray}\ttfamily,
	stringstyle=\color{red}\ttfamily,
	morestring=[b]',
	morestring=[b]"
}

\newcommand{\slither}{\textsc{Slither}\xspace}

\newcommand{\oyento}{\textsc{Oyente}\xspace}

\newcommand{\smartcheck}{\textsc{SmartCheck}\xspace}
\newcommand{\ourTool}{\textsc{Vulpedia}\xspace}
\newcommand{\mythril}{\textsc{Mythril}\xspace}

\newcommand{\zeus}{\textsc{Zeus}\xspace}
\newcommand{\vuddy}{\textsc{Vuddy}\xspace}
\newcommand{\securify}{\textsc{Securify}\xspace}
\definecolor{burnOrange}{cmyk}{0, 0.7808, 0.4429, 0.1412}
\newcommand{\todo}[1]{#1}
\definecolor{newpurple}{rgb}{0.58, 0, 0.33}
\newcommand{\keyw}[1]{\textcolor{blue}{#1}}
\newcommand{\gram}[1]{\textcolor{newpurple}{\textit{#1}}}

\def\WithComments{}
\ifdefined \WithComments

\newcommand{\hlt}[1]{\textcolor{red}{#1}}

\else

\newcommand{\hlt}[1]{\textcolor{black}{#1}}

\fi

\newcommand{\codeff}[1]{\texttt{\footnotesize #1}}

\newcommand{\totalContract}{76,354\xspace}
\newcommand{\testContract}{{17,770}\xspace}

\makeatletter
\newcommand{\linebreakand}{%
  \end{@IEEEauthorhalign}
  \hfill\mbox{}\par
  \mbox{}\hfill\begin{@IEEEauthorhalign}
}
\makeatother

\usepackage[nounderscore]{syntax}
\shortverb{\|}
\begin{document}

\title{\ourTool: Detecting Vulnerable Ethereum Smart Contracts via Abstracted Vulnerability Signatures}

\author{\IEEEauthorblockN{Jiaming Ye}
\IEEEauthorblockA{Kyushu University \\
Japan \\
ye.jiaming.852@s.kyushu-u.ac.jp
}
\and
\IEEEauthorblockN{Mingliang Ma}
\IEEEauthorblockA{Nanjing University \\
China}
\and
\IEEEauthorblockN{Yun Lin}
\IEEEauthorblockA{National University of Singapore \\
Singapore}
\linebreakand
\IEEEauthorblockN{Lei Ma}
\IEEEauthorblockA{University of Alberta \\
Canada}
\and
\IEEEauthorblockN{Yinxing Xue}
\IEEEauthorblockA{University of Science and 
Technology of China \\
China}
\and
\IEEEauthorblockN{Jianjun Zhao}
\IEEEauthorblockA{Kyushu University \\
Japan}
}

\maketitle

This article is accepted by Journal of Systems and Software.

\input{sec/abs}
\input{sec/1-intro}

\input{sec/2-background}

\input{sec/3-overview}
\input{sec/4-empirical}

\input{sec/5-learn_sig}

\input{sec/6-case-study}
\input{sec/7-evaluation}
\input{sec/8-related}
\input{sec/9-conclusion}

\input{sec/ref.tex}

\end{document}

%% file: sec/abs.tex
\begin{abstract}

Recent years have seen smart contracts are getting increasingly popular in building trustworthy decentralized applications.
Previous research has proposed static and dynamic techniques
to detect vulnerabilities in smart contracts. These tools check vulnerable contracts against several predefined rules. However, the emerging new vulnerable types and programming skills to prevent possible vulnerabilities emerging lead to a large number of false positive and false negative reports of tools. To address this, we propose \ourTool, which mines expressive vulnerability signatures from contracts. \ourTool is based on the relaxed assumption that the owner of contract is not malicious. Specifically, we extract structural program features from vulnerable and benign contracts as \textit{vulnerability signatures}, and construct a systematic detection method based on detection rules composed of vulnerability signatures. Compared with the rules defined by state-of-the-arts, our approach can extract more expressive rules to achieve better completeness (i.e., detection recall) and soundness (i.e., precision). 
We further evaluate \ourTool with four baselines (i.e., Slither, Securify, SmartCheck and Oyente) on the testing dataset consisting of \testContract contracts. The experiment results show that \ourTool achieves best performance of precision on 4 types of vulnerabilities and leading recall on 3 types of vulnerabilities meanwhile exhibiting the great efficiency performance. 
\end{abstract}

%% file: sec/1-intro.tex
\section{Introduction}

Powered by the Blockchain technique~\cite{BeckART17}, smart contracts~\cite{smartcontract} have attracted much attention and been applied in various industries, e.g., financial service, supply chains, smart traffic, and IoTs. Solidity is the most popular language for smart contract for its mature tool support and simplicity. However, the public has witnessed several severe security incidents, including the notorious DAO attack~\cite{daoAttack} and Parity wallet hack~\cite{parityHack}. According to previous reports~\cite{AtzeiBC16,sigmaprime}, up to 16 types of security vulnerabilities were found in Solidity programs. These security issues undermine the confidence of people who have executed transactions via smart contracts and eventually affect the trust towards the Blockchain ecosystem.

Witnessing the severity and urgency of this problem, researchers and security practitioners have made endeavors to develop automated security scanners~\cite{Git-Slither,Git-Oyente,zeus,securify,smartcheck}. Existing state-of-the-art scanners usually adopt the rule-based methods for vulnerability detection. \slither~\cite{Git-Slither} supports 39 hard-coded static rules; \securify~\cite{securify} supports 15 rules for verifying the extracted path constraints from the contract with the SMT solvers~\cite{datalog}; \oyento~~\cite{Git-Oyente} supports 8 rules for generating assertions for verifying the vulnerabilities. Each rule represents a pattern of vulnerable contract, which warns the programmers to avoid potential risks before deploying the contracts. 

Although experiments have demonstrated their effectiveness, it is notable that rules behind these scanners are manually crafted by human experts. \emph{The manually predefined rules can be obsolete}, because \ding{182} previously unseen vulnerable code may be introduced, which cannot be captured by the hard-coded rules, and \ding{183} new defense mechanisms (i.e., programming skills to prevent bugs) may have successfully mitigated the vulnerabilities, but the code may still match the predefined vulnerable pattern or rules. Therefore, most updated rules should be learned to distinguish vulnerable contracts from robust ones.

In this work, we alleviate the incompleteness of detection rules by combining vulnerability signatures abstracted from both vulnerable and benign contracts (i.e., vulnerable signature and benign signature). The vulnerable signature is designed for matching commonalities of a particular vulnerability. Comparatively, the benign signature is abstracted from falsely reported contracts in order to reduce false alarms. For each vulnerability, we adopt vulnerable and benign signatures to synthesize detection rules of \ourTool. Note that \ourTool is built upon the relaxed assumption that the owner of contract is not malicious. Detecting malicious contract (e.g., contract with backdoors, exploit code) is different to the vulnerability detection (i.e., the target of \ourTool). Based on this assumption, the operations related to the contract owner are all deemed as vulnerability defense behaviors. Compared with previous work, the synthesized rules are more updated and expressive than the predefined rules in the state-of-the-art vulnerability scanners, capturing a lot of unseen patterns in practice. 

In our implementation, we first collect truly and falsely reported contracts by applying three state-of-the-art vulnerability scanners (i.e., \slither~\cite{Git-Slither}, \oyento~\cite{oyento}, and \securify~\cite{securify}) and manually evaluate their correctness. Based on the results, analyzing truly reported vulnerable contracts allows us to capture salient program signatures responsible for vulnerable contracts. In contrast, analyzing falsely reported vulnerable contracts allows us to capture signatures noticeable for avoiding false alarms. Next, we categorize the contracts by their vulnerability types (e.g., Reentrancy, Unchecked Low-level-call, etc.) and alarm types (i.e., true or false alarm). For each category, we cluster the contracts based on their tree edit distance~\cite{PawlikA15} and then extract program feature commonalities from the PDGs (program dependency graph) of each cluster to summarize vulnerability signatures. Finally, we abstract \todo{4} vulnerable signatures and \todo{6} benign signatures. They are integrated as \todo{4} detection rules regarding \todo{4} vulnerabilities (i.e., Reentrancy, SelfDestruct, Tx-origin, and Unexpected-Revert). 

We conduct our signature abstraction on a set of \totalContract smart contracts and evaluations on a set of \testContract contracts, respectively. The evaluation results show that, compared with the state-of-the-art vulnerability scanners (i.e., \slither, \oyento, \smartcheck and \securify), our approach achieves outstanding accuracy on \todo{4} vulnerabilities and leading recall on \todo{3} vulnerabilities. We make our tool, \ourTool, available at \cite{mavs_link}.

To summarize, we make the following contributions:
\begin{enumerate}[leftmargin=*,topsep=0pt,itemsep=0ex]
\item We propose an approach to abstract vulnerability signatures and compose detection rules to report vulnerability. The learned rules are more expressive than rules of the state-of-the-art scanners, reporting vulnerabilities with better completeness and soundness
\item On the \testContract contracts crawled from Google, \ourTool yields the best precision on \todo{4} vulnerabilities and leading recall on \todo{3} ones, in comparison with the other state-of-the-art scanners.
\item Experiments show that \ourTool is efficient in vulnerability detection. The detection speed of \ourTool on \testContract contracts is far faster than \oyento and \securify.
\end{enumerate}

This work is organized as: In \autoref{sec:background}, we first introduce the different types of vulnerabilities we address in our study, and explain  why the state-of-the-art tools fail. In \autoref{sec:overview} we illustrate the basic steps of our proposed tool, namely \ourTool. In \autoref{sec:sig-abs}, we conduct an empirical study and introduce our method of signature abstraction. We also elaborate the effectiveness of signatures with examples. In \autoref{sec:expements}, we compare \ourTool with the other state-of-arts using \todo{17,770} real-world contracts deployed on Ethereum. \autoref{sec:related} briefly introduces the related work and \autoref{sec:conclusion} concludes our study.

%% file: sec/2-background.tex
\section{Background and Motivation} 
\label{sec:background}

In this section, we explain the \todo{4} vulnerability types (i.e., \codeff{Reentrancy}, \codeff{The abuse of tx.origin}, \codeff{Unexpected Revert} and \codeff{Self-destruct Abusing}.) targeted by our study. The 4 vulnerabilities deeply threats the safety of transactions of smart contracts. For example, the \codeff{Reentrancy} caused the DAO attack in 2016 and resulted in hundreds of millions dollars losses; The \codeff{tx.origin} and \codeff{Unexpected Revert} vulnerability are listed in the Decentralized Application Security Project (DASP)~\cite{dasp}; The \codeff{Self-destruct Abusing} vulnerability often appears with the use of \codeff{selfdestruct} instruction in Solidity, and is prone to being exploited if it is not well protected. We also show a real-world case, which is not well-handled by the state-of-the-art scanners, to motivate this work.

\subsection{Vulnerability Types} \label{sec:background:type}


\begin{enumerate}[leftmargin=*, topsep=0pt]
	\item \emph{Reentrancy (RE)} As the most famous Ethereum vulnerability, reentrancy recursively triggers the fall-back function~\cite{solidity-doc} to steal money from the victim's balance or deplete the gas of the victim. Reentrancy occurs when external callers manage to invoke the callee contract before the execution of the original call is finished, and it was mostly caused by the improper usages of the function \codeff{withdraw()} and \codeff{call.value(amount)()}. It was also reported in \cite{AtzeiBC16}.
	
	\item \emph{The Abuse of \codeff{tx.origin} (TX)} When the visibility is improperly set for some key functions (e.g., some sensitive functions with \codeff{public} modifier), the extra permission control then matters. However, issues can arise when contracts use the deprecated \codeff{tx.origin} (especially,  \codeff{tx.origin==owner}) to validate callers for permission control. It is relevant to the \emph{access control} vulnerability in \cite{dasp}. When a user \emph{U} calls a malicious contract \emph{A}, who intends to forward call to contract \emph{B}. Contract \emph{B} relies on vulnerable identity check (e.g., \codeff{require(tx.origin == owner)} to filter malicious access. Since \codeff{tx.orign} returns the address of \emph{U} (i.e., the address of \codeff{owner}), malicious contract \emph{A} successfully poses as \emph{U}.
	
	
	\item \emph{Unexpected Revert (UR)} In a smart contract, some operations may unfortunately fail. This can lead to two main impact: 1) the gas (i.e., the fee of executing an operation in Ethereum platform) of the transaction is wasted; 2) the transaction will be reverted, i.e., the denial of service (DoS). The denial of service attack is also termed "DoS with revert" in~\cite{consensys}. The attacker could deliberately make some operations fail for the purpose DoS.
	For example, some functions recursively send ethers to an array of users. If one of these calls fails, the whole transaction will be reverted. An attacker can deliberately fail this transaction to achieve a denial-of-service attack.
	
	\item \emph{Self-destruct Abusing (SD)} This vulnerability allows the attackers to forcibly send Ether without triggering its fall-back function. Normally, the contracts place important logic in the fall-back function or making calculations based on a contract's balance. However, this could be bypassed via the \codeff{self-destruct} contract method that allows a user to specify a beneficiary to send any excess ether \cite{consensys}. That is, a vulnerable contract is prone to being exploited to transfer all money to attacker' account meanwhile shut down the service.
	
\end{enumerate}


\subsection{Motivating Examples}\label{sec:motivating-example}

\autoref{fig:reentrancy_example} is mistakenly alarmed by \slither and \oyento. The function \texttt{withdraw} intends to send ethers to the \texttt{msg.sender}. It first verifies the identity of caller at line 2. Then, the function reads the amount of current balance of the caller at line 3 and sends ethers to the caller by using a Solidity call \texttt{.call.value()()}. Finally, the function updates the balance of caller at line 5.

The reason of the false alarm of \slither is due to that \slither detects reentrancy with the following rule:
\begin{mdframed}
[
linewidth = 1pt,
innertopmargin = 0pt,
innerleftmargin = 0pt,
outerlinewidth = 0pt,
rightmargin =0pt,
leftmargin = 0pt
]
\begin{equation} \label{rule:slither:reentrance}
\small
\begin{split}
\gram{DataDep}(\_,var_g) \succ \gram{Call}(\_,var_g) \succ \\
\gram{DataDep}(\_,var_g) \Rightarrow \text{reentrancy}
\end{split}
\end{equation}
\end{mdframed}

In \autoref{rule:slither:reentrance}, \gram{DataDep}(\_,$var\_g$) denotes write and read operations to variables; $var_{g}$ denotes a certain public global variable; $\succ$ denotes the execution order in the control flow; \gram{Call}(\_,$var_g$) denotes function call operations. This rule describes a common pattern for Reentrancy vulnerability. \autoref{fig:reentrancy_example} shows a typical example. $var_{g}$ is usually a balance account (e.g., \texttt{balances[msg.sender]}, line 3 in \autoref{fig:reentrancy_example}). An attacker just needs to create a fallback function that calls \texttt{withdraw()}. Once \texttt{msg.sender.call.value(amount)()} is executed and transfers the funds, the attacker’s fallback function~\cite{solidity-doc} will be triggered and call \texttt{withdraw()} (line 1) again. By this means, the attacker can transfer more funds before \texttt{balances[msg.sender]} is reduced to 0. This continues until there is no \textit{ether} remaining, or execution reaches the maximum stack size.

However, the pattern in \autoref{rule:slither:reentrance} is usually an over-estimation for real Reentrancy vulnerability. In fact, the example in \autoref{fig:reentrancy_example} is a counter-example because the function \texttt{withdraw()} is protected by an identity check at line 2. This statement specifies a precondition for running the \texttt{withdraw()} function. Once the precondition is not satisfied, the execution will be aborted. In \autoref{fig:reentrancy_example}, the identity check indicates that the contract calling this \texttt{withdraw()} function is limited to its \texttt{owner} (i.e., the creator of the contract).

\begin{figure}[t]
\begin{lstlisting}
function withdraw() {
    require(msg.sender == owner);
    uint256 amount = balances[msg.sender];
    require(msg.sender.call.value(amount)());
    balances[msg.sender] = 0;
}
\end{lstlisting}
\caption{An example of a non-vulnerable code. This is misreported as vulnerability by \slither and \oyento.}
\label{fig:reentrancy_example}
\end{figure}

The reason of the false alarm of \oyento is due to that \oyento detects reentrancy with the following rule:

\begin{mdframed}
[
linewidth = 1pt,
innertopmargin = 0pt,
innerleftmargin = 0pt,
outerlinewidth = 0pt,
rightmargin =0pt,
leftmargin = 0pt
]
\begin{equation} 
\label{rule:oyente:reentrance}
\small
\begin{split}
(\gram{DataDep}(\_,var_g) \wedge (gas_{trans} > 2300) \wedge \\
(amt_{bal} > amt_{trans})) \succ \gram{Call}(\_,var_g) \Rightarrow \text{reentrancy}
\end{split}
\end{equation}
\end{mdframed}

In \autoref{rule:oyente:reentrance}, \oyento requires the gas expense less than a certain value. In Solidity programs, each transaction requires an amount of gas to complete in the runtime. $gas_{trans} > 2300$ means the gas used for transaction must be larger than \todo{2300} (\todo{2300} is the least gas expense to conduct a transaction call). $amt_{bal} > amt_{trans}$ means the balance amount must be larger than transfer amount. Finally, the rule of \oyento
requires call to external functions by \gram{Call} meanwhile send money. Comparing with \autoref{rule:slither:reentrance} (defined by \slither), \oyento has more constraints for gas and balance value. Similar to the rule of \slither in  \autoref{rule:slither:reentrance}, \autoref{rule:oyente:reentrance} also overestimates the condition where Reentrancy attack can happen. With the protection by the identity check (i.e., line \todo{2} in \autoref{fig:reentrancy_example}), the execution of function calls conforms to the defined runtime conditions but is already free from the Reentrancy attack.

\noindent\textbf{How \ourTool can address this issue:} In contrast, \ourTool is equipped with detection rule composed of a vulnerable signature as shown in \autoref{rule:reen:vulnerble-sig} (i.e., the signature indicating potential vulnerability) and a benign signature as shown in \autoref{rule:reen:benign-sig} (i.e., the signature indicating potential code behaviors defending or fixing vulnerability). 

\begin{mdframed}
[
linewidth = 1pt,
innertopmargin = 0pt,
innerleftmargin = 0pt,
outerlinewidth = 0pt,
rightmargin =0pt,
leftmargin = 0pt
]
	\begin{equation} 
	\label{rule:reen:vulnerble-sig}
	\small
	\begin{split}
    \gram{DataDep}(\_,var_g) \succ \gram{Call}(\_,var_g) \\ \Rightarrow reentrancy
	\end{split}
	\end{equation}
\end{mdframed}

\begin{mdframed}
[
linewidth = 1pt,
innertopmargin = 0pt,
innerleftmargin = 0pt,
outerlinewidth = 0pt,
rightmargin =0pt,
leftmargin = 0pt
]
	\begin{equation} 
	\label{rule:reen:benign-sig}
	\small
	\begin{split}
    \gram{ControlDep}(\keyw{msg.sender},\_) \succ  
    \gram{DataDep}(\_,var_g) \succ \\ \gram{Call}(\_,var_g) \Rightarrow reentrancy
	\end{split}
	\end{equation}
\end{mdframed}

For the vulnerable signature, \ourTool adopts valuable experience from \slither and \oyento and detects Reentrancy by matching data dependency of variables followed by call operations. As for the benign signature, \ourTool eliminates false reports by filtering out functions which contain control dependency on \keyw{msg.sender}. For example, in \autoref{fig:reentrancy_example} the code at lines 2 checks if the \keyw{msg.sender} equals the address of owner, and the function is not considered as vulnerability by \ourTool.

%% file: sec/3-overview.tex
\section{Overview} 
\label{sec:overview}

\begin{figure*}[t]
	\centering  
	\includegraphics[scale=0.55]{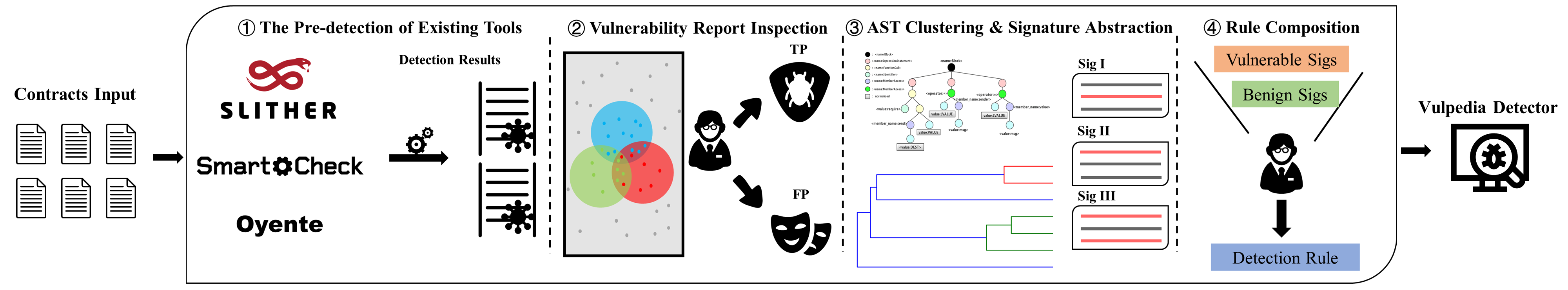}
	\caption{The workflow of extracting vulnerability signatures of \ourTool.}
	\label{fig:sys}
\end{figure*}




\autoref{fig:sys} shows the workflow of abstracting vulnerability signatures for \ourTool. The workflow can be roughly grouped into four steps: 1) The pre-detection of existing tools; 2) Vulnerability report inspection; 3) AST clustering and signature abstraction; 4) Rule composition. Note that manual efforts are involved in step 2 and step 4. 

In the first two steps, we systematically evaluate (1) how accurately state-of-the-art tools can report the vulnerable smart contracts and (2) under what condition can those tools be ineffective. We collect the reports of the state-of-the-art tools on a training dataset of \totalContract contracts. Then, we employ three experienced smart contract developers to manually confirm the reports of the tools, and categorize them into two groups: truly alarmed vulnerable contracts and falsely alarmed vulnerable contracts.

In the last two steps, we first calculate the tree edit distance based on the ASTs of contracts in a particular vulnerability type and cluster the contracts of the type by defining the contract similarity. Next, we abstract common nodes from the PDGs (program dependency graph) of each cluster to summarize signatures (e.g., as shown in \autoref{fig:learningSignature}). From truly vulnerable contracts, we summarize vulnerable signatures. In contrast, from falsely alarmed vulnerable contracts, we summarize benign signatures. Finally, we manually integrate the vulnerable signatures and benign ones into vulnerability detection rules.

\begin{figure}[t]
	\centering  
	\includegraphics[scale=0.4]{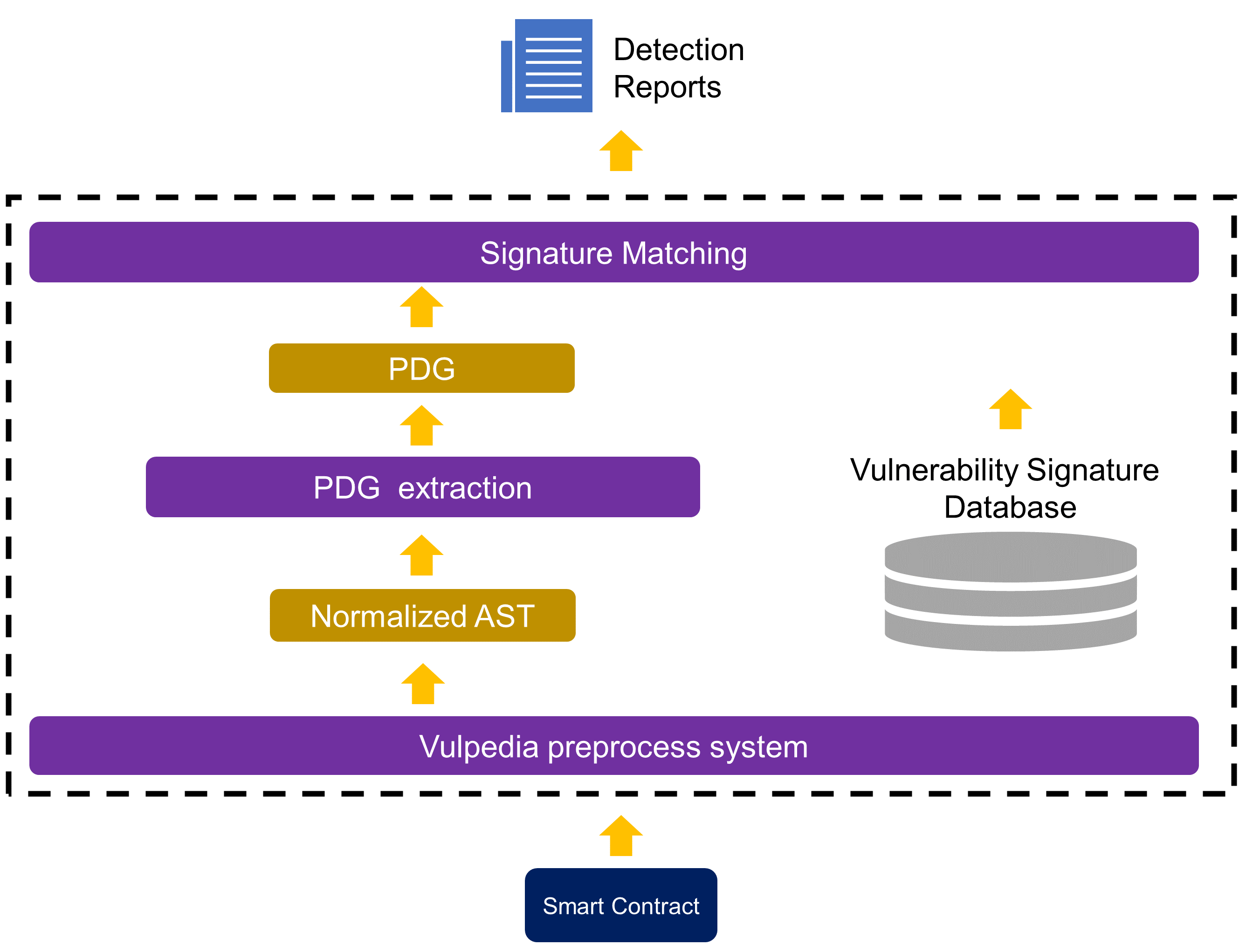}
	\caption{The architecture diagram of Vulpedia.}
	\label{fig:arche}
\end{figure}

After we equip our \ourTool detector with the composed rules, the detector takes unknown contracts as inputs and generate vulnerability reports based on the signatures. The figure of the architecture is shown in \autoref{fig:arche}. Specifically, the detector first conduct preprocessing on the input smart contract code. The detector extracts normalized AST from the contract. Based on this normalized AST, the detector conducts a PDG extraction. Meanwhile, the detector extracts existing signatures from the vulnerability signature database. Lastly, the detector produces detection reports based on the comparison results. That is, if the PDG matches vulnerable signatures but not matched with benign signatures, the contract will be deemed as a vulnerable contract; otherwise, the detector will produce a non-vulnerable report.

%% file: sec/4-empirical.tex
\section{Empirical Study of Signature Abstraction} 
\label{sec:sig-abs}

In this section, we first illustrate how we empirically collect contracts in this study. We report how we select the vulnerability scanners and how we construct a contract dataset. Next, we introduce our method of \ding{182} clustering similar contracts by comparing tree edit distance, \ding{183} abstracting commonalities from PDGs of clusters as signatures and \ding{184} detection rules composition based on the abstracted signatures. Finally, we elaborate the signatures with examples to evidence the representativeness of them.

\subsection{Selected Scanners and Dataset}

\begin{table*}[h]
    \centering
    \caption{The state-of-art tools for Solidity analysis.}
    \small
    \begin{tabular}{cccccc}
\toprule
 Tool Name & Method & Technique & Open Source & Implementation & Adopted In Experiment\\
\midrule
 \textsc{Mythril}~\cite{Git-Mythril}  & Dynamic & Constraint Solving &  \ding{108} & Python & \ding{109}\\
 \textsc{MythX}~\cite{Git-Mythx}  & Dynamic & Constraint Solving & \ding{109} &  N.A.  & \ding{109}\\
 \textsc{Slither}~\cite{Git-Slither} & Static & CFG Analysis & \ding{108} & Python & \ding{108}\\
 \textsc{Echidna}~\cite{Git-Echidna}  & Dynamic & Fuzzy Testing & \ding{108} & Haskell & \ding{109}\\
 \textsc{Manticore}~\cite{Git-Manticore}  & Dynamic & Testing & \ding{108} & Python & \ding{109}\\
 \textsc{Oyente}~\cite{oyento} & Dynamic & Constraint Solving & \ding{108} & Python & \ding{108}\\
 \textsc{SmartCheck}~\cite{smartcheck}  & Static & AST Analysis & \ding{108} & Java & \ding{108}\\
 \textsc{Octopus}~\cite{Git-Octopus}  & Static & Reverse Analysis & \ding{108} & Python & \ding{109}\\
 \textsc{Zeus}~\cite{zeus} & Static & Formal Verification & \ding{109} & N.A. & \ding{109}\\
 \textsc{ContractFuzzer}~\cite{contractfuzzer} & Dynamic & Fuzzy Testing & \ding{108} & Go & \ding{109}\\
\bottomrule
\end{tabular}
\label{tab:all_tools}
\end{table*}

\subsubsection{Choice of Scanners and Vulnerability Types}
Overall, we select vulnerability scanners based on how practical they can be used in real-world scenarios. 
We investigate a list of static analyzers, including  \slither~\cite{Git-Slither}, \oyento~\cite{oyento}, \zeus~\cite{zeus} \textsc{SmartCheck}~\cite{smartcheck}, and \textsc{MythX}~\cite{Git-Mythx}. These tools utilize manually defined detection rules to detect vulnerabilities. The rules could match vulnerabilities in some cases but also generate much false reports. We also investigate dynamic detectors like \textsc{Mythril}~\cite{Git-Mythril}, \textsc{ContractFuzz}~\cite{contractfuzzer}, \textsc{Echidna}~\cite{Git-Echidna} and \textsc{Manticore}~\cite{Git-Manticore}. They exercise programs and check the runtime status of functions to find vulnerabilities. The dynamic analyzers often achieve high detection precision but suffer from limited scalability. Additionally, we investigate other analyzing tools (e.g., Solidity reverse engineering tool \textsc{Octopus}~\cite{Git-Octopus}) to facilitate our exploiting contracts. A summary of the above tools can be found at \autoref{tab:all_tools}. In our study, some tools are not selected because they are not open-sourced (\zeus~\cite{zeus}, \textsc{MythX}~\cite{Git-Mythx}), not related to our task (\textsc{Echidna}~\cite{Git-Echidna}, \textsc{Octopus}~\cite{Git-Octopus}) and efficiency concerns (\textsc{Mythril}~\cite{Git-Mythril}, \textsc{ContractFuzzer}~\cite{contractfuzzer}, \textsc{Manticore}~\cite{Git-Manticore}).

Finally, we choose \slither v.0.4.0, \oyento  v0.2.7 and \smartcheck v2.0 as our scanners.

\subsubsection{Dataset for Empirical Study}

We implement a web crawler to download Solidity files from accounts of Etherscan~\cite{etherscan}, a famous third-party website on Ethereum block explorer. Etherscan provides APIs for downloading transaction information (e.g., transaction addresses, time).

\begin{table}[h]
    \centering
    \caption{The percentages of adopted Solidity contracts versions. According to~\cite{solidityversion}.}
    \begin{tabular}{cccc}
    \toprule
    Major Version & \# of Smart Contracts & Percentage \\
    \midrule
     0.1 & 13 & $<$0.1\% \\
     0.2 & 89 & $<$0.1\% \\
     0.3 & 519 & 0.39\% \\
     0.4 & 71,350 & \textbf{54.27\%} \\
     0.5 & 32,479 &  24.69\% \\
     0.6 & 22,171 & 16.85\% \\
     0.7 & 4,200 & 3.19\% \\
     0.8 & 725 & 0.55\% \\
    \bottomrule
    \end{tabular}
    \label{tab:solidity versions}
\end{table}

We choose contracts deployed by Solidity 0.4.25 and 0.4.24. The reasons are two folds: 1) as listed in \autoref{tab:solidity versions}, Solidity 0.4 is the majority version among all versions and the 0.4.24 and 0.4.25 are the latest versions in Solidity 0.4; 2) The versions 0.4.24 and 0.4.25 are supported by most analyzers, so that they facilitate our study. Additionally, we find that the downloaded dataset has redundant contracts (contracts which share commonality with others). Regarding these redundant contracts, we remove contracts that are exactly same to others and contracts that are only different in transfer address with others. 
Finally, we have \totalContract contracts for empirical study.
Our crawler can be accessed at \url{https://github.com/ToolmanInside/smart_contract_crawler}.

\autoref{tab:empirical_contracts} shows the number of contracts we collected in this study. Overall, among \totalContract contracts, the three tools report \todo{508} true vulnerable contracts albeit \todo{3,496} false vulnerable ones. \autoref{tab:empirical_tools} shows the details on the number of reported contracts and precision performance of each tool. We observed that all the tools have a large number of false alarms. This is due to contract programmers have invented many heuristics to detect the potential vulnerabilities. In other words, most existing detection rules are obsolete. It motivates us to pursue (and generate) a more expressive and fine-grained rule to mitigate the false alarms.

\begin{table}[t]
\centering
\small
\caption{Number of collected contracts for each category}
\label{tab:empirical_contracts}
\begin{tabular}{ccccc}
\toprule
\textbf{Alarm Type} & \textbf{RE} & \textbf{TX} &  \textbf{UR} & \textbf{SD} \\ 
\midrule
True Positive  & 46 & 38  & 421  & 3 \\ 
False Positive & 720 & 179 & 2,546 & 51 \\ 
\bottomrule
\end{tabular}
\end{table}

\begin{table}[t]
\centering
\small
\tabcolsep=0.05cm
\caption{The precision performance of three tools \slither, \oyento and \smartcheck on four vulnerabilities.}
\label{tab:empirical_tools}
 \begin{tabular}{cccc}
 \toprule
  \textbf{Vulnerability} & \slither & \oyento & \smartcheck  \\
  \midrule
  RE & 623 (3.53\%) & 143 (16.78\%) & N.A.  \\
  TX & 67 (28.35\%) & N.A. & 150 (12.66\%) \\
  UR & 2,678 (8.25\%) & N.A. & 289 (69.20\%) \\
  SD & 54 (5.56\%) & N.A. & N.A.  \\
\bottomrule
 \end{tabular}%
 \label{tab:old_rq1}%
\end{table}%

%% file: sec/5-learn_sig.tex
 \begin{figure*}
  \centering
  \includegraphics[width=\linewidth]{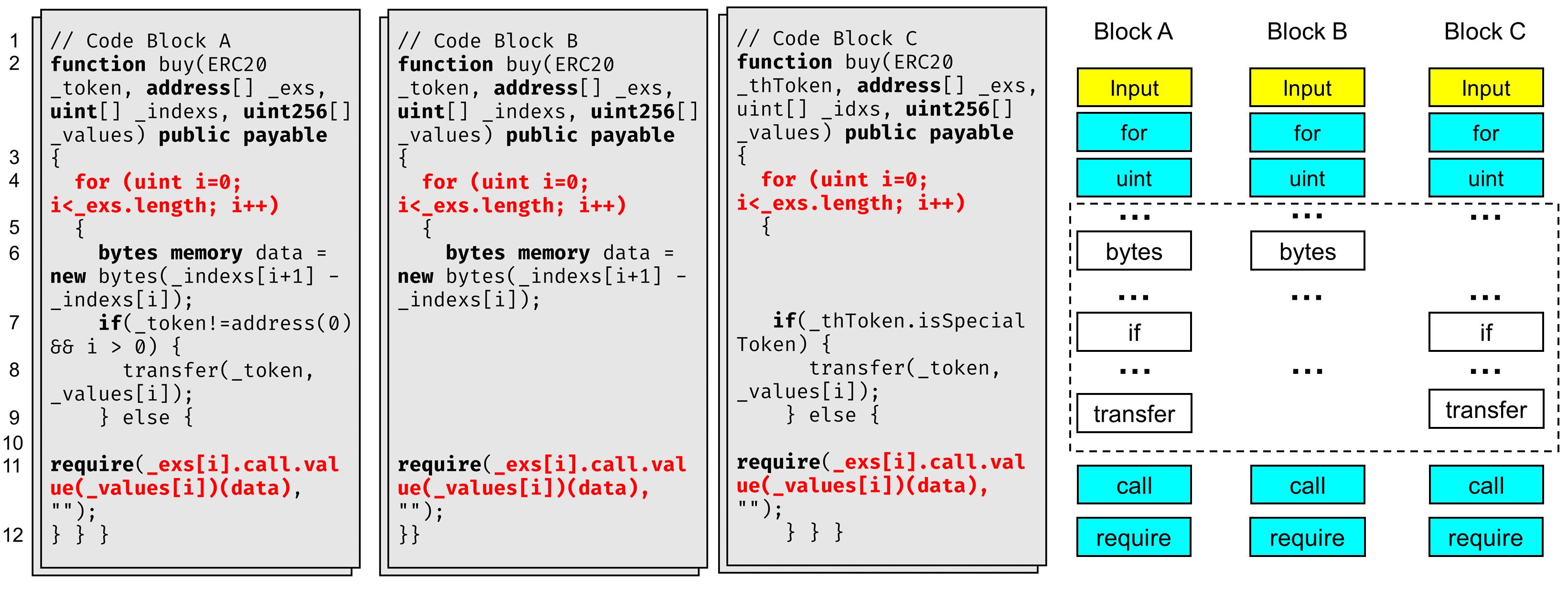}
  \caption{Three similar code blocks of Unexpected Revert that are found in real-world contracts. Based on their tree edit distance, we cluster them together and abstract a graph skeleton from their PDG. The yellow boxes denote function inputs, blue boxes denote common nodes on PDG and white boxes in dotted box represent different nodes.}
  \label{fig:learningSignature}
 \end{figure*}

\subsection{Vulnerability Rule Abstraction} 
\label{sec:AVS}
In this section, we introduce the definition of signature and show how we cluster and abstract the vulnerable/benign signatures from each cluster.

\subsubsection{Definition}
We define a vulnerability rule for a Solidity contract as following BNF:

\begin{center}
  \begin{minipage}{\linewidth}
    \begin{grammar}
<rule> ::= <comp\_sig> 

<comp\_sig> ::= $\lnot$<comp\_sig> | (<comp\_sig> $\lor$ <comp\_sig>) | (<comp\_sig> $\wedge$ <comp\_sig>) | \\
(<comp\_sig> $\succ$ <comp\_sig>) | <sig>

<sig> ::= \gram{DataDep}(\textit{X},\textit{Y}) | \gram{ControlDep}\textit{(X,Y)} | \gram{ForLoop} | \\
\gram{IsInstance}\textit{(X,Y)} | \gram{Call}\textit{(L,X)} | \gram{SelfDestruct}(\textit{X}) | \keyw{msg.sender} | \keyw{tx.origin} | 
    \end{grammar}
  \end{minipage}
\end{center}

Here, the detection rule is composite of signatures. A composite signature is a negation ($\lnot$) of itself, or conjunction ($\wedge$), union ($\lor$), succeed ($\succ$) with another composite signature. A composite signature can also be a single vulnerability signature. Specifically, vulnerability signature indicates basic program relationships and built-in keywords of Solidity language. For example, the data dependency (\gram{DataDep}(\textit{X},\textit{Y})) relationship denotes that variable \textit{X} has data dependency to \textit{Y} (i.e., variable assignment operations). The control dependency (\gram{ControlDep}\textit{(X,Y)}) indicates assertation operations (e.g., require, assert) between variables \textit{X} and \textit{Y}. The for loop \gram{ForLoop} denotes the function body exists a for loop statement. The \gram{IsInstance}\textit{(X,Y)} denotes the variable \textit{X} is a type of variable \textit{Y}. Call operation \gram{CALL}\textit{(L,X)} includes low level calls (e.g., \texttt{call.value()} and \texttt{send()} in Solidity) and high level calls (i.e., user-defined function calls). Here, variable \textit{L} represents the result of call operations and variable \textit{X} represents the parameters required by the call. \gram{SelfDesutrct}(\textit{X}) is a built-in function call in Solidity. Once it is called, the service of current contract is stopped and the rest balance is transferred to an arbitrary receiver \textit{X}. The \keyw{msg.sender} and \keyw{tx.origin} are built-in variables. Specifically, \keyw{msg.sender} denotes the address of current contract and \keyw{tx.origin} denotes the origin of call chains~\cite{solidity-doc}.

\begin{algorithm}[t]
	\small
	\SetKwInOut{Input}{input}
	\SetKwInOut{Output}{output}
    \Input{$SourceCode$, source code of smart contracts}
	\Output{$SignatureCands$, abstracted signature candidates}
    // \texttt{Contract Clustering Process} \\
    ASTs = $getAST(SourceCode)$ \\
    nASTs = $ASTNomalization(ASTs)$ \\
    distanceMatrix = $List[N,N]$ \\
    // \texttt{N is the number of trees} \\
    \ForEach{idx $i \in range(nASTs)$}{
        \ForEach {idx $j \in range(nASTs) and i \neq j$}{
            treeEdtDist = ARTED(nASTs[i], nASTs[j]) \\
            // \texttt{Calculate the distance between two trees} \\
            distanceMatrix[i, j] = treeEdtDist \\
        }
    }
    Clusters = $hierarchicalClustering(distanceMatrix)$ \\

    // \texttt{Signature Abstraction Process} \\
    $SignatureCands \leftarrow \emptyset$ \\
    \ForEach{cluster $c \in Clusters$}{
        $PDGs \leftarrow \emptyset$ \\
        \ForEach{tree $t \in c$}{
            PDG $p \leftarrow getPDG(t)$ \\
            $pn \leftarrow PDGNormalization(p)$ \\
            $PDGs \leftarrow PDGs \cup pn$ \\
        }
        $commonSeq \leftarrow LCS(PDGs)$ \\
        $SignatureCands \leftarrow SignatureCands \cup commonSeq$
    }
    \textbf{return} $SignatureCands$
	\caption{Contract Clustering and Signature Abstraction Algorithm}\label{algo:clustering}
\end{algorithm}

\subsubsection{Contract Clustering} In this section, we first define contract similarity on normalized ASTs, and then we cluster similar trees by using hierarchical clustering algorithm. The clustering procedure can be found at line 1 to line 10 in the \autoref{algo:clustering}.

\noindent\textbf{Contract Similarity.} We define the contract similarity by considering both semantic and structural information of the code. To this end, we use AST (Abstract Syntax Tree) to represent the code of the functions of each contract.
For each AST of a Solidity function, we normalize the concrete nodes in the AST for retaining core information and abstracting away unimportant details such as variable names or constant values, as shown in line 2 of \autoref{algo:clustering}. For each AST corresponding to a function, we just retain the information such as node type, name, parameter and return value (if contained). For the variable names (e.g., \codeff{\_indexs} in code block A and code block B of \autoref{fig:learningSignature}), they will be normalized with the token asterisk  \lq\lq{$\ast$}\rq\rq{}. Similarly, we repeat the same normalization for constant values of the types  \codeff{string},  \codeff{int},  \codeff{bytes} or  \codeff{uint}.

Given two trees, we use the tree edit distance between two normalized ASTs as their distance. The AST normalization process is shown in line 3 of \autoref{algo:clustering}. In this work, we apply a robust algorithm for the tree edit distance (ARTED)~\cite{PawlikA15}, which computes the optimal path strategy by performing an exhaustive search in the space of all possible path strategies. Here, \emph{path strategy} refers to a mapping between two paths of the two input trees (or subtrees), as the distance between two (sub)trees is the minimum distance of four smaller problems, i.e., (1) the edit distance between two empty trees, (2) the edit distance of transferring a tree \textit{F} to an empty tree, (3) the edit distance of transferring an empty tree  to a tree \textit{F} and (4) the edit distance of transferring a tree \textit{F} to another tree \textit{G}. Note that though ARTED runs in \emph{quadratic} time and space complexity, it is guaranteed to perform as good or better than its competitors~\cite{PawlikA15}.

\noindent\textbf{Contract Clustering.} We cluster the ASTs via hierarchical clustering algorithm with complete linkage~\cite{completeLink}, as shown in line 4 to line 10 in \autoref{algo:clustering}. Then, we group the codes in \autoref{fig:learningSignature} with considerable modification. We deem that the ASTs in each cluster share commonalities as a feature (or signature) for a vulnerability category.

\subsubsection{Signature Abstraction}
\label{sec: signature-abstraction}
After clustering contract functions with AST, we abstract signature by referring to their PDG (Program Dependency Graph) information. The reason lies in that PDG allows us to capture the code semantic features like control and data dependencies.

\noindent\textbf{PDG Representation.}
For each AST, we transfer its code into a PDG including all its depended code elements such as global variables and called functions, as shown in line 13 to line 17 in \autoref{algo:clustering}. In a PDG, each of its nodes is an instruction and the edge between nodes indicates data dependency, control dependency, and call relation between the nodes. Thus, given a cluster containing $N$ Solidity functions, we reduce it into a problem of finding the common subgraph of $N$ PDGs. The normalization of PDGs is shown in line 18 in \autoref{algo:clustering}.

\noindent\textbf{PDG Matching.}
The graph matching problem is a NP-complete problem. We simplify the problem with the following steps. Before matching, we also abstract away variable names and constant values in the PDGs as we do that for AST. Next, we simplify the calculation by flattening the graph into a node sequence (via depth first order search) and align the sequences by LCS algorithm~\cite{lcs}, as shown in line 20 in \autoref{algo:clustering}.
The aligned graph nodes are considered as commonalities shared by the code in the same cluster.



As a result, the signature abstracted from a cluster is essentially a graph skeleton, as shown in \autoref{fig:learningSignature}. Then, we manually inspect those skeletons and refine them into usable signatures. The refining process requires manual efforts because some signatures are semantically similar to others but different in syntax. These signatures require to be filtered out by human expert.
After we repeat the above procedures on both vulnerable and benign contracts, we construct a set of vulnerable and benign signatures.

\begin{table*}[h]
\centering
\normalsize
\caption{Extracted Signatures from Different Vulnerability Categories}
\label{tab:signatures}
\begin{tabular}{cccl}
\toprule
\textbf{ID} & \textbf{Vulnerability} & \textbf{V/B} & \textbf{Signature} \\
\midrule
1 & \multirow{4}{*}{Reentrancy} & V & \gram{DataDep}\textit{(\_,X)} $\succ$ \gram{Call}\textit{(\_,X)}\\
2 &  & B &  \gram{ControlDep}(\keyw{msg.sender},\textit{X}) $\succ$ \gram{DataDep}\textit{(\_,X)} $\succ$ \gram{Call}\textit{(\_,X)} \\
3 &  & B & \gram{DataDep}\textit{(\_,X)} $\succ$ \gram{IsInstance}(\textit{X},\keyw{addr}) $\succ$ \gram{Call}\textit{(\_,X)} \\
4 &  & B & \gram{ControlDep}\textit{(Y,\_)} $\succ$ \gram{DataDep}\textit{(\_,X)} $\succ$ \gram{Call}\textit{(\_,X)} $\succ$ \gram{DataDep}\textit{(Y,\_)} \\
\midrule
5 & \multirow{2}{*}{Unexpected Revert} & V & \gram{ForLoop} $\succ$ \gram{Call}\textit{(L,X)} $\succ$ \gram{ControlDep}\textit{(L,\_)} \\
6 & & B & \gram{ForLoop} $\succ$ \big(\gram{IsInstance}(\textit{X},\keyw{addr}) $\wedge$ \gram{Call}\textit{(L,X)} \big) $\succ$ \gram{ControlDep}\textit{(L,\_)} \\
\midrule
7 & \multirow{2}{*}{Abuse of Tx.origin} & V & \gram{DataDep}(\keyw{tx.origin},\textit{X}) $\succ$ \gram{ControlDep}\textit{(X,\_)} \\
8 & & B & \gram{DataDep}(\keyw{msg.sender},\textit{Y}) $\succ$ \gram{DataDep}(\keyw{tx.origin},\textit{X}) $\succ$ \gram{ControlDep}\textit{(X,Y)} \\
\midrule
9 & \multirow{2}{*}{SelfDestruct} & V & \gram{DataDep}\textit{(\_,X)} $\succ$  \gram{SelfDestruct}(X)\\
10 &  & B & \gram{ControlDep}\textit{(\keyw{msg.sender},\textit{X})} $\succ$ \gram{DataDep}\textit{(\_,X)} $\succ$  \gram{SelfDestruct}(X) \\
\bottomrule
\end{tabular}
\end{table*}

\subsubsection{Rule Composition}
\label{sec: rule-composition}
In this work, we follow the following heuristics to integrate the signatures into a rule. Generally, a \textit{detection rule is a composite boolean expression of vulnerability signatures}. Given a vulnerability category, a detection rule first requires the input contract match with the vulnerable signatures. The vulnerable signatures are essential ingredients of forming a vulnerability. Therefore, if the input contract is not matched with vulnerable signatures, the contract should be considered as invulnerable. Next, the input contract is required not to match with benign signatures. The benign signatures are the best practices to defend vulnerabilities. If the input contract matches with them, it suggests that the contract is capable for defending vulnerabilities and should not be reported as vulnerability. 

\begin{table}[t]
\small
\caption{Detection rules for each vulnerability}
\centering
\label{tbl:rule-composition}
\begin{tabular}{ccc}
\toprule
ID & Vulnerability & Rule \\
\midrule
1  & Reentrancy    & SIG1 $\wedge$ $\lnot$ (SIG2 $\vee$ SIG3 $\vee$ SIG4) \\
2  & Revert        & SIG5 $\wedge$ $\lnot$ SIG6 \\
3  & Tx.origin     & SIG7 $\wedge$ $\lnot$ SIG8 \\
4  & Self-destruct & SIG9 $\wedge$ $\lnot$ SIG10 \\
\bottomrule
\end{tabular}
\end{table}

%% file: sec/6-case-study.tex
\subsection{Case Study: Abstracted Signatures}
\label{sec:casestudy}
We applied the three chosen scanners to \totalContract contracts. Overall, \slither~reports the most vulnerabilities, in total \todo{3,422} (623 + 67 + 2,678 + 54) candidates covering four types. In contrast, \smartcheck~reports \todo{439} (\todo{150} + \todo{289}) candidates and \oyento~reports only \todo{143} candidates. After they are processed by our methods, we abstract \todo{4} vulnerable signatures and \todo{6} benign signatures, as shown in \autoref{tab:signatures}. Based on these signatures, we further integrate them into \todo{4} detection rules, as shown in \autoref{tbl:rule-composition}. In this section, we elaborate the signatures with examples to evidence their representativeness.

\noindent\textbf{Signature of Reentrancy.} We extract \todo{4} signatures from TPs and FPs of reported reentrancy vulnerabilities, including \todo{1} vulnerable signature (\textbf{SIG1}) and \todo{3} benign signatures (\textbf{SIG2}, \textbf{SIG3}, \textbf{SIG4}). 

\textbf{SIG1} is abstracted from general patterns of reentrancy vulnerabilities. This signature consists of two parts: (1) the read or write operation of variable \textit{X} (i.e., \gram{DataDep}(\_,\textit{X})) and (2) the call operation with the parameter variable \textit{X} (i.e., \gram{Call}(\_,\textit{X})). 

\textbf{SIG2} adds various forms of checks (i.e., in \codeff{require} or \codeff{assert})  for \keyw{msg.sender} compared with SIG1. For example, SIG2 checks whether the identity of \keyw{msg.sender} is checked under certain conditions (e.g., equal to the owner, or with a good reputation, or having the dealing history) before calling the external payment functions. With the identity check, the function is only accessible to related users, blocking the malicious attack from attackers. Example of this signature can be found at \autoref{fig:reentrancy_example}.

\textbf{SIG3} describes a falsely reported case of transferring balance to a fixed address. In Fig.~\ref{fig:evaluation:fp1}, the function \codeff{closePosition} sends balance to a token \codeff{bancorToken} which is assigned with a fixed address at line 2. According to the detection rule of \slither (See \autoref{rule:slither:reentrance}), this code is a vulnerability because --- (1) it reads the \codeff{public} variable \codeff{agets[\_idx]}; (2) then calls external function \codeff{bancorToken.transfer()}; (3) last, writes to the \codeff{public} variable  \codeff{agets[\_idx]}. However, in practice, this contract can never be easily exploited to steal ethers due to the hard-coded address constant (i.e., \codeff{0x1F...FF1C}). Note that the constant address can be a malicious address, under such circumstance this address cannot protect contract. However, this case is very rare. Therefore, we choose to trust the creator of the contract as well as the designated addresses are benign.


\begin{figure}[t]
\begin{lstlisting}
contract BancorLender {
  ERC constant public bancorToken =
    ERC(0x1f573d6fb3f13d689ff844b4ce37794d79a7ff1c);
  function closePosition(uint _idx) public {
    ...
    bancorToken.transfer(agets[_idx].lender, amount);
    return;
    } }
\end{lstlisting}
\caption{A real case of using SIG3 (a hard-coded address at line 3), a FP of \emph{reentrancy} for \slither.}
\label{fig:evaluation:fp1}
\end{figure}

\begin{figure}[t] 
\begin{lstlisting}
contract ZethrBankroll is ERC223Receiving {
ZTHInterface public ZTHTKN;
bool internal reEntered;
function receiveDividends() public payable {
  if (!reEntered) {
    ...
    if (ActualBalance > 0.01 ether) {
      reEntered = true;
      ZTHTKN.buyAndSetDivPercentage.value(ActualBalance)(address(0x0), 33, "");
    }
} } }
\end{lstlisting}
\caption{A real case of using SIG4 (an execution lock of \codeff{reEntered}), an FP of \emph{reentrancy} for \slither.}
\label{fig:evaluation:fp4}
\end{figure}

\textbf{SIG4} is to prevent from the recursive entrance of the function --- eliminating the issue from root. For instance, in \autoref{fig:evaluation:fp4}, the internal instance variable \codeff{reEntered} will be checked at line 5 before processing the business logic between line 8 and 10. To prevent the reentering due to calling \codeff{ZTHTKN.buyAndSetDivPercentage.value()}, \codeff{reEntered} will be switched to \codeff{true}; after the transaction is done, it will be reverted to \codeff{false} to allow other transactions.



\noindent\textbf{Signature of Unexpected Revert.} We extract \todo{2} signatures from reported Unexpected Revert vulnerabilities, including \todo{1} vulnerable signature \textbf{SIG5} and \todo{1} benign signature \textbf{SIG6}.

\textbf{SIG5} represents general patterns of Unexpected Revert vulnerabilities. This signature consists of three parts: (1) the for loop program structure (i.e., \gram{ForLoop}); (2) the call operation of the variable \textit{X} (i.e., \gram{Call}(\_,\textit{X})); (3) the result of call operation is further checked by assertions. 

According to the recent technical article~\cite{nvestlabs2019}, the rules of \emph{Call/Transaction in Loop} are neither sound nor complete to cover most of the unexpected revert cases. At least, modifier \codeff{require} is often ignored, which makes \slither~and \smartcheck~incapable to check possible revert operations on multiple account addresses. Here,  multiple accounts must be involved for exploiting this attack --- the failure on one account blocks other accounts via reverting the operations for the whole loop. Hence, in the example of \autoref{fig:evaluation:fp6}, the operations in the loop are all on the same account (i.e., \codeff{sender} at line 5) and potential revert will not affect other accounts. Therefore, the transfer operation of which target is a single address is considered as \textbf{SIG6}.

\begin{figure}[t]
	\begin{lstlisting}
function withdraw() private {
  for(uint i = 0; i < player_[uid].planCount; i++) {
    ...
    address sender = msg.sender;
    sender.transfer(amount);
  }  }
	\end{lstlisting}
	\caption{A real FP of Unexpected Revert reported by \smartcheck, where only one account is involved (SIG6).}
	\label{fig:evaluation:fp6} 
\end{figure}

\noindent\textbf{Signatures of \texttt{Tx.Origin} Abusing.} We extract \todo{2} signatures from the truly vulnerable contracts and falsely reported contracts, including \todo{1} vulnerable signature (\textbf{SIG7}) and \todo{3} benign signatures (\textbf{SIG8}).

For \textbf{SIG7}, this signature is extracted from general patterns of tx.origin vulnerabilities. This vulnerability first reads the value of tx.origin, followed by an assignment to variable \textit{X} (i.e., \gram{DataDep}(\keyw{tx.origin},\textit{X})). After this, the function has an assertion to this variable (i.e., \gram{ControlDep}(\textit{X},\_)). While we extract signatures from the TPs of vulnerabilities, we find that our \textbf{SIG7} is slightly looser than the detection rule in \slither. \slither skips the function if there exists a read operation to a particular variable \keyw{msg.sender}, ignoring that some of these variables are irrelevant to \keyw{tx.origin}. In order not to overlook potential vulnerabilities, our \textbf{SIG7} only requires a read of \keyw{tx.origin}, followed by an assertion on this variable.



For \textbf{SIG8}, we observe that \smartcheck~reports much more cases (\todo{210}) than \slither (\todo{34}), but has lower precision performance than \slither. After our investigation, we find that the incorrect reports of \smartcheck are due to the unsound rules (as shown in \autoref{rule:smartcheck: tx.origin}). That is, \smartcheck simply reports vulnerability once \keyw{tx.origin} appears in assertion statements. However, under some circumstance (e.g., comparing \keyw{msg.sender} with \keyw{tx.origin}), the use of \keyw{tx.origin} should not be reported. We summarize the \textbf{SIG8} based on the FPs of \smartcheck.

\begin{mdframed}[
	linewidth = 1pt,
innertopmargin = 0pt,
innerleftmargin = 0pt,
outerlinewidth = 0pt,
rightmargin =0pt,
leftmargin=-5pt
]
	\begin{equation} 
	\label{rule:smartcheck: tx.origin}
	\small
\begin{split}
\gram{DataDep}(\keyw{tx.Origin},\textit{X}) \succ \gram{ControlDep}(\textit{X},\_)  \\
\Rightarrow \text{Tx.Origin abusing}
\end{split}
	\end{equation}
\end{mdframed}

\begin{figure}[t]
\begin{lstlisting}
function destroyDeed() public {
  require(msg.sender == owner);
  if (owner.send(address(this).balance)) {
    selfdestruct(burn);}
}
\end{lstlisting}
\caption{A real FP of \emph{self-destruct abusing} by \slither,  as \codeff{selfdestruct()} is used under two checks at line 2,3 (SIG10).}
\label{fig:evaluation:fp7}
\end{figure} 

\noindent\textbf{Signature of Self-destruct Abusing.} We extract \todo{2} signatures from the self-destruct vulnerabilities, including \todo{1} vulnerable signature \textbf{SIG9} and \todo{1} benign signature \textbf{SIG10}.

\textbf{SIG9} is extracted from general patterns of self-destruct vulnerabilities. This signature consists of two parts: (1) the read or write operation of variable \textit{X} (i.e., \gram{DataDep}(\_,\textit{X})) and (2) the call operation of the self-destruct with the parameter \textit{X} (i.e., \gram{SelfDestruct}(\textit{X})).

For \textbf{SIG10}, we extract this signature from FPs of tools. In the existing scanners, only \slither~detects the misuse of self-destruct, which is called suicidal detection. In total, \slither~reports \todo{54} cases of suicidal via its built-in rule --- as long as function \gram{SelfDestruct} is used, no matter what the context is, \slither~will report it. Obviously, the \slither's rule is too simple and too general. It mainly works for directly calling \gram{SelfDestruct} without permission control or conditions of business logic --- under such circumstance (\todo{3} out of \todo{54}), the \slither~rule can help to detect the abusing. In practice, in most cases (\todo{51} out of \todo{54}) \gram{SelfDestruct} is called with the \codeff{admin} or \codeff{owner} permission control or under some strict conditions in business logic. For example, \gram{SelfDestruct} is indeed required in the business logic at line 2 of \autoref{fig:evaluation:fp7}. As the owner wants to stop the service of the contract via calling \gram{SelfDestruct}, after the transactions are all done,  the contract becomes inactive. Note that parameter \codeff{burn} is just padded to call \gram{SelfDestruct} in a correct way. Hence, we summarize the \textbf{SIG10}, adding a strict condition control or a self-defined modifier for identity check when using \gram{SelfDestruct}.

In brief, for a vulnerability type, we use the vulnerable signatures to match potential vulnerabilities, which yield a better recall. Then, we leverage corresponding benign signatures to filter out false reports. 




\subsection{Vulnerability Detection}

\begin{figure}[t]
	\centering  
	\includegraphics[scale=0.4]{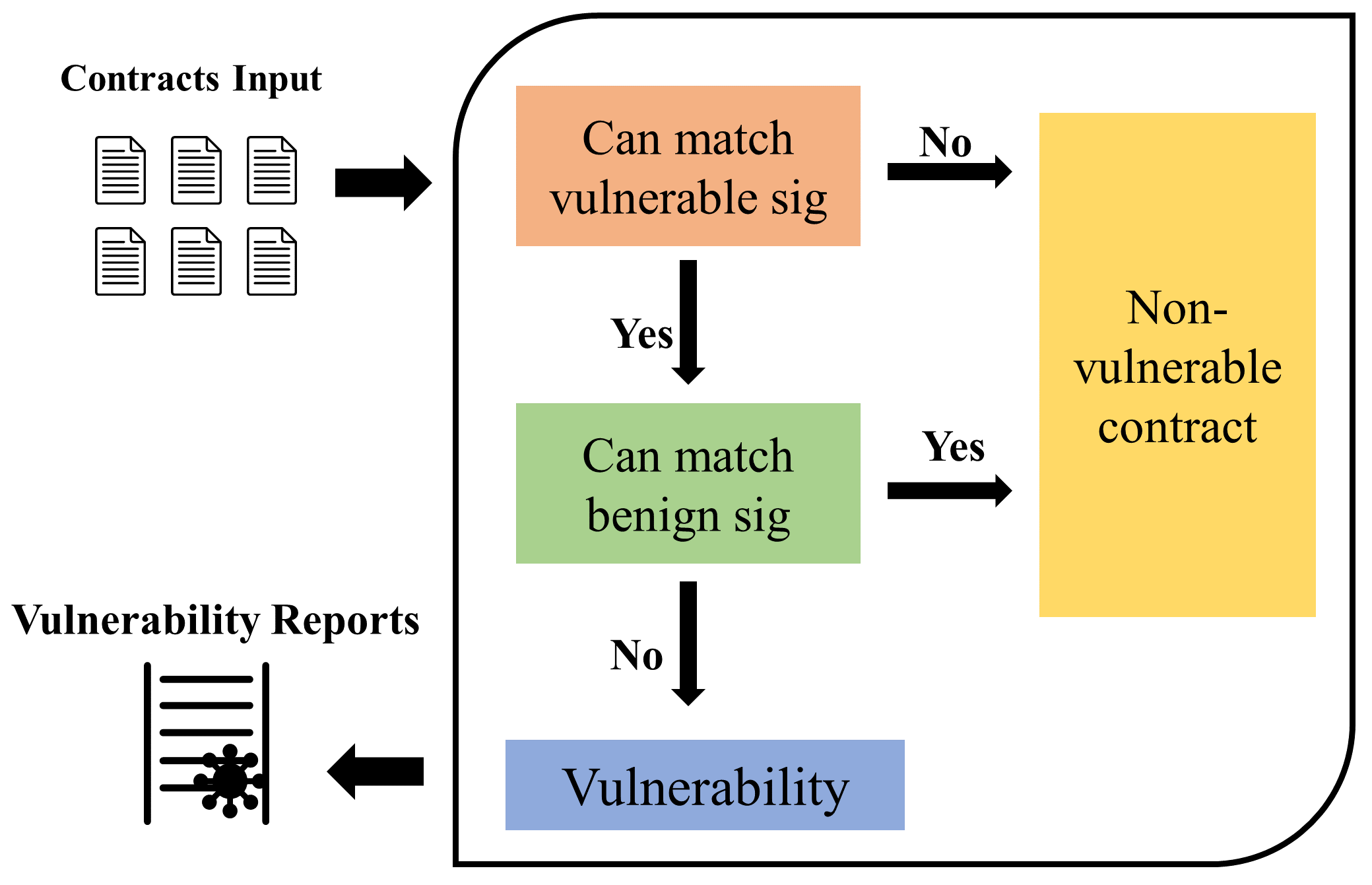}
	\caption{The detection workflow of Vulpedia.}
	\label{fig:detectionprocecss}
\end{figure}

The implementation of the vulnerability detection of \ourTool is based on the previously abstracted signatures and integrated detection rules, but slightly different from them. The workflow of detection is shown in \autoref{fig:detectionprocecss}. Specifically, in this workflow, \ourTool reports vulnerability only when the vulnerable signatures are matched meanwhile the benign signatures are not matched. That is, the vulnerable and benign signatures are separated things. However, in previous subsection, the signatures are combined to form detection rules. The reason is that our benign signatures are designed to filter out false positive reports. The detection rules shown in \autoref{tbl:rule-composition} all follow the pattern that the vulnerable signatures should be matched but the benign ones should not. Therefore, though the implementation of the detection process seems differently, the logic of the workflow is the same with previous designs.

%% file: sec/7-evaluation.tex
\section{Evaluation} \label{sec:expements}

\input{tables/detection_evaluation.tex}

\noindent\textbf{Experimental Environment.} Throughout the evaluation, all the steps are conducted on a machine running on Ubuntu 18.04, with 8 core 2.10GHz Intel Xeon E5-2620V4 processor, 32 GB RAM, and 4 TB HDD. For the scanners used in evaluation, no  multithreading options are available and only the  by-default setting is used for them. 

\noindent\textbf{Tool Implementation.} Vulpedia is implemented based on the \textsc{Slither} analyzer. We adopt the AST analysis from \textsc{Slither}, and we build PDG analysis based on the CFG (control flow graph) and call graph of \textsc{Slither}. The vulnerability signatures are implemented as detectors in nearly 1,000 lines of Python code. The demo of our tool can be found at \url{https://github.com/ToolmanInside/vulpedia_demo}.

\noindent\textbf{Dataset for Tool Evaluation.} To take a different dataset from contracts we used in empirical study, we get another address list of contracts from Google BigQuery Open Dataset. After removing contracts that already used in our empirical study, we get the other \testContract real-world contracts deployed on Ethereum, on which we fairly compare our resulted tool \ourTool with the version of the scanners: \slither~v0.6.4.  \oyento~v0.2.7,  \smartcheck~v2.0 and \securify~v1.0 that is open-sourced at Dec 2018. The evaluation dataset is opened along with empirical study dataset at \url{https://drive.google.com/file/d/1kizsz0_8B8nP4UNVr0gYjaj25VVZMO8C}.

The evaluations are conducted based on a relaxed assumption that the owners of contracts are not malicious. That is, the operations of the owners are all deemed as defense behaviors to vulnerabilities. The evaluations aim to answer these RQs:
\begin{enumerate}[leftmargin=*,label=\textbf{RQ\arabic*.},topsep=0pt,itemsep=0ex]
\item How is the precision of \ourTool, compared with the existing scanners in vulnerability detection?
\item How is the recall of \ourTool? Can our signature-based method report more vulnerabilities?
\item How is the efficiency of \ourTool, in tool comparison on the datasets?
\end{enumerate}

\subsection{RQ1: Evaluating the Precision of Tools} \label{sec:expements:otherRst}



As mentioned in \autoref{sec:casestudy}, we have learned \todo{10} signatures in total for the four types of vulnerabilities. To evaluate the effectiveness of the resulted vulnerable signatures and detection rules, we apply them on the \testContract newly collected contracts and compare with the other state-of-the-arts detection tools. Details on the performance of each tool are shown in Table~\ref{tab:rq1}. Note that all TPs are manually verified by our authors.

In Table~\ref{tab:rq1}, we list \todo{280} detection results of \ourTool, with an averaged precision of \todo{50.2\%}, regardless of vulnerability types. In comparison, \slither has an averaged precision of \todo{18.9\%}; \oyento's averaged precision is \todo{7.1\%}; \smartcheck's averaged precision is  {40.2\%}; and \securify's precision is surprisingly only \todo{1.1\%}. In the rest of this section, we analyze the false positives of these tools from the perspective of supporting vulnerability signatures.

\noindent\textbf{FPs of Reentrancy.} Among the four supported tools except \smartcheck, \ourTool yields the lowest FP rate (\todo{71.5\%}) owing to the adoptions of benign signatures for reentrancy. FP rates of other tools are even higher. For example, the FP rate of \securify is \todo{98.9\%}, as its detection pattern is too general but has not considered possible defense to vulnerabilities in code. \slither adopts Rule \ref{rule:slither:reentrance} to detect, but it supports no benign signatures --- its recall is acceptable, but FP rate is high. \oyento adopts Rule \ref{rule:oyente:reentrance} and has no benign signatures --- its recall is low due to the strict rule, and its FP rate is also high.

\noindent\textbf{FPs of Unexpected Revert.}  As summarized in \autoref{sec:casestudy}, \slither reports Unexpected Revert vulnerability when a call in loop is detected, ignoring the potential false alarms (i.e., low level call in a loop). This coarse detection rule leads to \todo{335} FPs. \smartcheck handles \todo{SIG5} but not \todo{SIG6} and leads to \todo{27} FPs. In comparison, \ourTool combines \textbf{SIG5} and \textbf{SIG6} for integrating detection rule, yielding the lowest FP rate \todo{51.2\%}.

\noindent\textbf{FPs of \texttt{Tx.Origin} Abusing.} \slither has a strict rule for detecting this type, only checking the existence of \codeff{tx.Origin == msg.sender}. We find that this tool also skips the function if there exists a read operation to a particular variable \keyw{msg.sender}, ignoring that some of these variables are irrelevant to \keyw{tx.origin}. For the case that \keyw{tx.origin} is compared with an unrelated address variable, \slither reports it as vulnerability, causing FPs. Comparatively, \smartcheck and \ourTool manage to include all the identity check cases, but meanwhile also lead to FPs due to the fact --- accurate symbolic analysis is not adopted in \smartcheck or \ourTool to suggest whether \keyw{tx.Origin} can be used to rightly replace \keyw{msg.sender}. Hence, the FP rate due to ignoring \textbf{SIG8} is higher than that of \ourTool.

\noindent\textbf{FPs of Self-destruct Vulnerability.} \ourTool has \todo{13} FPs. After inspecting, we find \todo{10} FPs are due to the unsatisfactory handling of \textbf{SIG10}. That is, the identity check hides in self-defined modifiers. Function modifiers are overlooked by \ourTool, causing FPs. Comparatively, \slither only reports \todo{3} true positives. The reason is that \slither simply reports vulnerability when a \gram{SelfDestruct} call is detected. Due to the inconsideration of the potential access controls, \slither performs less precision than \ourTool.

\begin{tcolorbox}[size=title]
  {\textbf{Answer to RQ1:} \ourTool performs best in evaluations of precision among tools. In detecting \codeff{tx.origin} vulnerability, \ourTool outperforms the second best tool by \todo{45.3\%} (88.7\% - 44.3\%). The reason of the high precision performance is \ourTool adopts effective benign signatures to remove false reports.
  }
\end{tcolorbox}

\subsection{RQ2: Evaluating the Recall of Tools} 

\begin{figure}
  \centering
  \includegraphics[width=\linewidth]{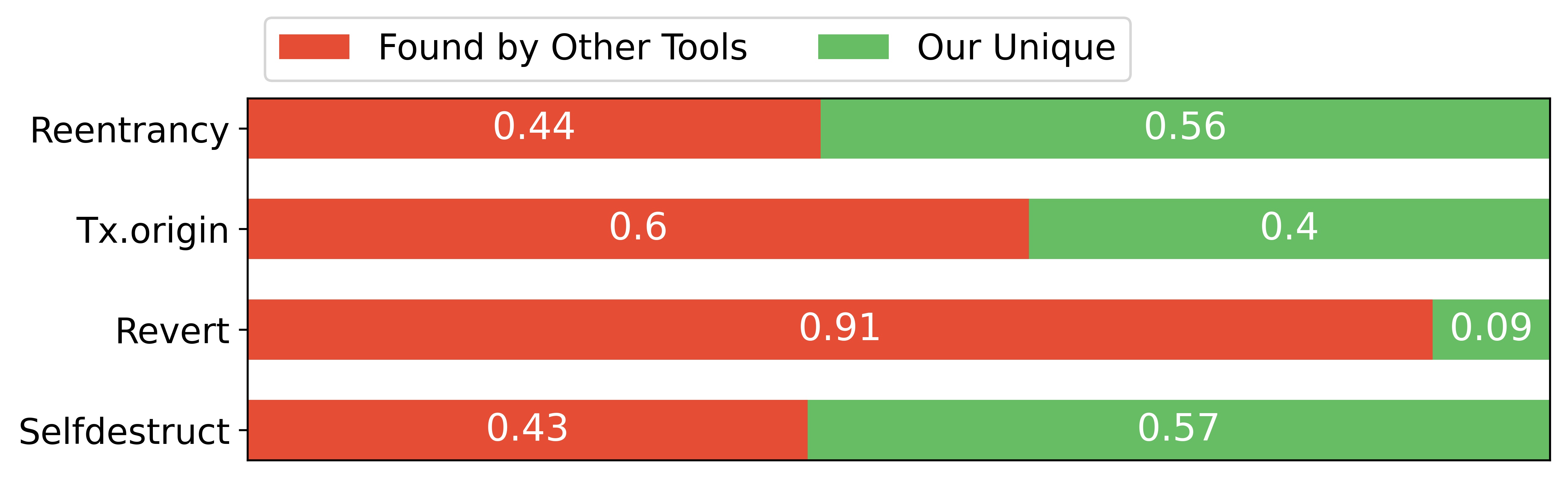}
  \caption{Comparing the vulnerabilities only reported by \ourTool with vulnerabilities reported by other tools. ``Our Unique'' means those only found by \ourTool.}
  \label{fig:unique-tp}
 \end{figure}

In \autoref{tab:rq1}, in most cases, \ourTool yields the best recall except on unexpected revert, where R\% for \smartcheck is {77.4}\% and R\% for \ourTool is {67.7}\%.
Based on the vulnerable signature abstracted in empirical study, we expect \ourTool can find more similar vulnerable candidates. A comparison between vulnerabilities \emph{only} found by \ourTool (denoted by green bars) and vulnerabilities found by other tools (represented by red bars) is shown in \autoref{fig:unique-tp}. 



\noindent\textbf{Recall of Reentrancy.} In this vulnerability, \ourTool performs best by report \todo{69.3\%} vulnerabilities. Among all TPs, \ourTool finds \todo{56\%} unique TPs that are missed by other evaluated tools. We find that the other three tools commonly fail to consider the user-defined function \codeff{transfer()}, not the built-in payment function \codeff{transfer()}. For the example in Fig.~\ref{fig:evaluation:fn1}, \slither and \securify miss it as they mainly check the external call for low-level functions (e.g., \codeff{send(), value()}) and built-in \codeff{transfer()}, ignoring user defined calls. \oyento does not report this example, as it fails in the balance check according to Rule~\ref{rule:oyente:reentrance}. Comparatively, \ourTool detects this vulnerability, as we have an vulnerable signature that has a high code similarity with this example. Notably, though \ourTool has the best recall of \todo{69.3\%}, it misses \todo{30.7\%} TPs. This is due to the fact that reentrancy has many forms, and our vulnerable signature is not sufficient to cover those TPs.  

\noindent\textbf{Recall of Unexpected Revert.} In this vulnerability, the performance of \ourTool is slightly worse than \smartcheck (77.4\%). Specifically, \ourTool only reports \todo{9\%} unique TPs while \todo{91\%} TPs are found by other tools. 
The reason of the TPs missed by \ourTool (reported by \smartcheck) are due to the incompleteness of our vulnerable signature \textbf{SIG6}. That is, the signature requires a \gram{ControlDep} after \gram{Call}. However, the \gram{ControlDep} is unnecessary when the \gram{Call} is a high level call (e.g., user defined function call) because assertion operations are already integrated in high level calls. Therefore, the signature causes FNs.

\noindent\textbf{Recall of \texttt{\small Tx.Origin} Abusing.} For this type, \todo{96.6\%} TPs are found by \ourTool --- almost all TPs are found by \ourTool. Additionally, \ourTool reports \todo{40\%} unique TPs which are missed by other tools. The reason is that we matches identity check of \codeff{Tx.Origin} in self-defined modifiers, which is commonly overlooked by other tools. 


\noindent\textbf{Recall of Self-destruct Abusing.} For this type, all vulnerabilities (100\%) are found by \ourTool. Comparatively, \slither only reports \todo{42.8\%} vulnerabilities. \todo{57\%} of TPs are only found by \ourTool. The rationale of TPs missed by \slither is that \slither skips the function if the function is only accessible to internal calls (i.e., set visibility to \codeff{internal}). These functions are however prone to being exploited by internal calls. Therefore, they should not be overlooked. \ourTool leverages \textbf{SIG9} to match vulnerability candidates, so we have better recall performance.

\begin{tcolorbox}[size=title]
  {\textbf{Answer to RQ2:} \ourTool has best performance on detection recall. Except Unexpected Revert, \ourTool outperforms other tools on three vulnerabilities. The reason of this leading performance is our abstracted vulnerable signatures can represent essence of most vulnerabilities. 
  }
\end{tcolorbox}

\input{fig/eval_fn1.tex}
\subsection{RQ3: Evaluating the Efficiency} \label{sec:expements:comparison}

\begin{table}[t]
\centering
\caption{The time (min.) of vulnerability detection for each scanner on 76,354 and 17,770 contracts. ``S.C.'' denotes \smartcheck.}
\label{tab:efficiency}
\begin{tabular}{cccccc}
\toprule
Dataset & \slither & \oyento & \textsc{S.C.} & \securify & \ourTool \\
\midrule
76,354  & 156     & 6,434   & 641        & N.A.     & 883      \\
17,770  & 52      & 1,352   & 141        & 8,859     & 295 \\
\bottomrule
\end{tabular}
\end{table}

\noindent\textbf{On Dataset for Empirical Study.}  
In Table \ref{tab:efficiency}, \slither takes the least time (only \todo{156} \emph{min}) in detection. \smartcheck and \ourTool have the comparable detection time (500$\sim$1000 \emph{min}). They are essentially of the same type of technique --- pattern based static analysis. In practice, they may differ in performance due to implementation differences, but still, they are significantly faster than \oyento that applies symbolic execution. Compared with other dynamic analysis or verification tools (i.e., \textsc{Mythrill} and \textsc{Securify} that cannot finish in three days for the \totalContract contracts), \oyento is quite efficient. Notably, the signature abstraction time of \ourTool is not included in the detection time, as it could be done off-line separately. Since signature abstraction is analogical to rules formulation, it is not counted in the detection time. 

\noindent\textbf{On Dataset for Tool Evaluation.} On the smaller dataset,  we observe the similar pattern of time execution --- \slither is the most efficient, \oyento is least efficient (except \securify), and \smartcheck and \ourTool have the comparable efficiency. Notably,  \securify can finish the detection on \testContract contracts, but it takes significantly more time than other tools. The performance issue of \securify rises due to the conversion of EVM IRs into datalog representation and then the application of verification technique. \oyento is also less efficient, as it relies on symbolic execution for analysis. \ourTool should be comparable to \smartcheck and \slither, as these three all adopt rule based matching analysis. The extra overheads of \ourTool, compared with \slither and \smartcheck, are signature-based code matching.

\begin{tcolorbox}[size=title]
  {\textbf{Answer to RQ3:} \ourTool outperforms \securify and \oyento regarding the detection efficiency on both empirical evaluation and tool comparison. In general, \ourTool is efficient as a signature-based vulnerability detection tool.}
\end{tcolorbox}

\subsection{Threats to Validity}

In our experiments, we adopt recall rate as a metric, which is a potential threat. Generally, the recall rate indicates the number of TPs divided by the number of all vulnerabilities. However, it requires an overwhelming effort to find out all vulnerabilities (i.e., the ground truth). In our study, we evaluate recall performance based on the union of vulnerabilities reported by all tools. Additionally, in the abstraction of signatures, we manually confirm signatures, which may introduce bias. To alleviate this, we repeat our experiments for \todo{3} times. Also, we note that the randomness is an inevitable factor in the evaluations of efficiency. We repeat the experiments for \todo{5} times and record the average values. Besides, the abstracted signatures are prone to introducing incompleteness. To alleviate this, we implement our methods on the top of \slither, which facilitates our signature abstraction from PDGs.

%% file: tables/detection_evaluation.tex
\begin{table*}[htbp]
	\footnotesize
	\centering
	\caption{The detection performance for our tool and other existing ones on the \testContract contracts, where \#N refers to the number of detections, P\% and R\% refer to the precision rate and the recall rate among the number of detections, respectively. Note that P\%= (\#TP of the tool)/\#N,  and R\%= (\#TP of the tool)/ (\#TP in union of all tools).}
	\begin{tabular}{cccccccccccccccc}
		\toprule
		Vulnerability & \multicolumn{3}{c}{\slither} & \multicolumn{3}{c}{\oyento} & \multicolumn{3}{c}{\smartcheck} & \multicolumn{3}{c}{\securify} & \multicolumn{3}{c}{\ourTool} \\ \cline{2-16}
		          & \#N & P\% & R\% & \#N & P\% & R\% & \#N &P\% & R\% & \#N & P\% & R\% & \#N & P\% & R\% \\
		\midrule
		Reentrancy & 162   & 9.8\% & 32.6\%  & 28 & 7.1\% & 4.1\% & N.A. & N.A. & N.A. & 797 & 1.1\% & 18.3\% & 119  & \textbf{28.5\%}  & \textbf{69.3\%} \\
		Abuse of tx.origin & 23  & 43.4\% & 33.3\%  & N.A. & N.A. & N.A. & 44 & 33.3\% & 56.6\% & N.A. & N.A. & N.A. & 98    & \textbf{88.7\%}  & \textbf{96.6\%} \\
		Unexpected Revert & 356   & 5.8\%  & 67.7\%  & N.A. & N.A. & N.A. & 51 & 47.1\% & \textbf{77.4\%} & N.A. & N.A. & N.A. & 43    & \textbf{48.8\%}  & 67.7\% \\
		Self Destruct & 18    & 16.6\%  & 42.8\%    & N.A. & N.A. & N.A. & N.A. & N.A. & N.A. & N.A. & N.A. & N.A. & 20    & \textbf{35.0\%}  & \textbf{100\%} \\
		\bottomrule
	\end{tabular}%
	\label{tab:rq1}%
\end{table*}%

%% file: fig/eval_fn1.tex
\begin{figure}[t]
	\vspace{-4mm}
\begin{lstlisting}
contract Alice {
  ...
  function aliceClaimsPayment(bytes32 _dId, uint _amount, address _addr) external {
    require(deals[_dId].state==DS.Initialized);
    ...
    deals[_dId].state = DS.PaymentSentToAlice;
    if (_addr == 0x0) {msg.sender.transfer(_amount);}
    else {
      ERC20 token = ERC20(_addr);
      assert(token.transfer(msg.sender, _amount)); }
  }
}
\end{lstlisting}
\caption{A real case of reentrancy. This is a TP for \ourTool but a FN for \slither, \oyento and \securify.}
\label{fig:evaluation:fn1}
\end{figure}

%% file: sec/8-related.tex
\section{Discussions}

\subsection{The Relaxed Security Assumption}

The experiments and comparisons are all conducted based on the relaxed security assumption. That is, we assume the operations of contract owner are not malicious behaviors. We follow this assumption because the security-design is more strict than ordinary contract when the contract is designed for industry needs. In fact, existing successful contracts (e.g., e-voting, NFT) have been audited by experts to be protected from rogue owners. To avoid our tool been blindly used by users and developers, this assumption should be pointed out.

\subsection{The Weakness of \ourTool}

In this section, we discuss the improvement of the weakness of \ourTool found in our experiment practice. In our view, involving manual efforts brings biases, and the biases may affect the effectiveness of the tool. However, \ourTool relies on manual efforts, mainly in the two steps: 1) manually confirm the reports of existing tools in our empirical study. We add man-powers in this step because the existing static tools have severe limitations and produce a large number of false reports. Due to Ren et al.~\cite{issta2021smartcontract}, the \slither tool has a false positive rate over 70\%. If the false reports are not removed from all reports, the dataset cannot be correctly labeled, and our signature abstraction is infeasible. 2) We manually integrate the vulnerable signatures and benign ones into vulnerability detection rules. In this step, we use manual efforts to filter out ineffective signatures. This is due to the lack of smart contract vulnerability benchmark. If we have a benchmark, we can replace the man-powers in this step and filter out ineffective signatures by running testing on the benchmark.


\section{Related Work} \label{sec:related}


\noindent\textbf{Vulnerability Detection in Smart Contracts.} There is already a list of security scanners on smart contracts. From the perspective of software analysis, these scanners could be categorized into static- or dynamic-based. In the former category, \slither~\cite{Git-Slither} aims to be the analysis framework that runs a suite of vulnerability detectors.
\textsc{Oyente}~\cite{oyento} analyzes the bytecode of the contracts and applies Z3-solver~\cite{z3} to conduct symbolic executions. Recently, \smartcheck~\cite{smartcheck} translates Solidity source code into an XML-based IR and defines the XPath-based patterns to  find code issues. \textsc{Securify}~\cite{securify} is proposed to detect the vulnerability via compliance (or violation) patterns to guarantee that certain behaviors are safe (or unsafe, respectively).   These static tools  usually adopt symbolic execution or verification techniques, being relevant to \ourTool. However, none of them applies code-similarity based matching technique or takes into account the possible DMs in code to prevent from attacks. 

There are some other tools that enable the static analysis for smart contracts. \textsc{VeriSmart}~\cite{verismart} proposes a domain-specific algorithm for verifying smart contracts. \textsc{VerX}~\cite{verx} combines symbolic execution and contract status abstraction to verify transactions. \zeus \cite{zeus} adopts  XACML as a language to write the safety and fairness properties, converts them into LLVM IR~\cite{llvm_ir} and then feeds them to a verification engine such as \textsc{SeaHorn} \cite{GurfinkelKKN15}. Besides, there is another EVM bytecode decompiling and analysis frame, namely \textsc{Octopus}~\cite{Git-Octopus}, which needs the users to define the patterns for vulnerability detections. To prevent the DAO, Grossman et al. propose the notion of effectively Callback Free (ECF) objects in order to allow callbacks without preventing modular reasoning \cite{ecf}. \textsc{Maian} is presented to detect greedy, prodigal, and suicidal contracts \cite{maian}, and hence the vulnerabilities to address differ from the types we address in this paper. The above tools are relevant, but due to various reasons (e.g., issues in tool availability), we cannot have a direct comparison with them. 
 
The less relevant category includes dynamic testing or fuzzing tools:  \textsc{Manticore}~\cite{Git-Manticore}, \mythril~\cite{Git-Mythril}, \textsc{MythX}~\cite{Git-Mythx}, \textsc{Echidna}~\cite{Git-Echidna} and \textsc{Ethracer}~\cite{ethracer}.  \textsc{sFuzz}~\cite{sfuzz} and \textsc{Harvey}~\cite{harvey} use the advanced techniques (e.g., concolic testing, fuzzing and tainting) for detection. Dynamic tools often target certain vulnerability types and produce results with few FPs. However, they are unsuitable for a large-scale detection due to the efficiency issue.  

\noindent\textbf{Code-similarity based Vulnerability Detection.} In general, similar-code matching technique is widely adopted for vulnerability detection. In 2016, \textsc{VulPecker}~\cite{VulPecker} is proposed to apply different code-similarity algorithms in various purposes for different vulnerability types. It leverages vulnerability signatures from  National Vulnerability Database (NVD)~\cite{NVD} and applies them to detect 40 vulnerabilities that are not published in NVD, among which 18 are zero-days. As \textsc{VulPecker} works on the source code of C,   \textsc{Bingo}~\cite{bingo} can execute on binary code and compare the assembly code via tracelet (partial trace of CFG) extraction~\cite{tracy} and similarity measuring. \vuddy~\cite{vuddy}  targets at exact clones and parameterized clones, not gapped clones, as it utilizes hashing for matching for the purpose of high efficiency.
To sum up, these studies usually resort to the vulnerability database of C language for discovering similar zero-days. In contrast, plenty of our efforts are exhausted in gathering vulnerabilities from other tools for smart contracts and auditing them manually. \ourTool adopts a more robust algorithm (e.g., LCS), which can tolerate big or small code gaps across the similar candidates of a vulnerability. 
 

%% file: sec/9-conclusion.tex
\section{Conclusion}
\label{sec:conclusion}

In this study, we propose \ourTool, a static analyzer based on abstracted signatures. We focus on addressing one key challenges: the manually predefined detection rules can be obsolete. To this end, we first conduct an empirical study for signature abstraction. We leverage state-of-the-arts scanners to detect vulnerabilities on our training dataset. Based on their results, we propose a method to cluster similar contracts and abstract vulnerable signatures and benign signatures, respectively. After we collect all signatures, we conduct comparative evaluations with state-of-the-art tools. The results show that \ourTool exhibits best performance of precision on \todo{4} types of vulnerabilities and leading recall on \todo{3} types of vulnerabilities with great efficiency performance.

%% file: sec/ref.tex
\bibliographystyle{IEEEtran}
\bibliography{ref}